\documentclass[onecolumn]{aa}  
\usepackage{verbatim}
\usepackage{color}

\newcommand       \cm         {\,{\rm cm}}
\newcommand       \mum      {\,{\rm \mu m}}

\newcommand       \s            {\,{\rm s}}
\newcommand       \ppm       {\,{\rm ppm}}
\newcommand       \eV          {\,{\rm eV}}

\newcommand     \gtsim  {\lower.5ex\hbox{$\buildrel > \over \sim$}}
\newcommand     \ltsim  {\lower.5ex\hbox{$\buildrel < \over \sim$}}
\newcommand     \simgt  {\lower.5ex\hbox{$\buildrel > \over \sim$}}
\newcommand     \simlt  {\lower.5ex\hbox{$\buildrel < \over \sim$}}
\newcommand       \simali       {{\sim\,}}
\begin{document}

\title{Synthesizing carbon nanotubes in space}

   \author{Tao Chen
          \inst{1}\fnmsep\thanks{Email: taochen@kth.se}
          \and
          Aigen Li
          \inst{2}\fnmsep\thanks{Email: lia@missouri.edu}
          }
          
   \institute{
         School of Engineering Sciences in Chemistry, Biotechnology and Health, Department of Theoretical Chemistry \& Biology, Royal Institute of Technology, 10691, Stockholm, Sweden
         \and
         Department of Physics and Astronomy, University of Missouri, Columbia, MO 65211, USA
          }

\abstract
  {As the 4th most abundant element in the universe, carbon (C) is widespread in the interstellar medium (ISM) in various allotropic forms (e.g., fullerenes have been identified unambiguously in many astronomical environments, the presence of polycyclic aromatic hydrocarbon molecules in space has been commonly admitted, and presolar graphite as well as nanodiamonds have been identified in meteorites). As stable allotropes of these species, whether carbon nanotubes (CNTs) and their hydrogenated counterparts are also present in the ISM or not is unknown.}
     {We explore the possible routes for the formation of CNTs in the ISM and calculate their fingerprint vibrational spectral features in the infrared (IR).}
     {We study the hydrogen-abstraction/acetylene-addition (HACA) mechanism and investigate the synthesis of nanotubes using density functional theory (DFT). The IR vibrational spectra of CNTs and hydrogenated nanotubes (HNTs), as well as their cations, have also been obtained with DFT.} 
     {We find that CNTs could be synthesized in space through a feasible formation pathway. CNTs and cationic CNTs, as well as their hydrogenated counterparts, exhibit intense vibrational transitions in the IR. Their possible presence in the ISM could be investigated by comparing the calculated vibrational spectra with astronomical observations made by the {\it Infrared Space Observatory}, {\it Spitzer Space Telescope}, and particularly the upcoming {\it James Webb Space Telescope}.}
     {}


   \maketitle

\section{Introduction}
Carbon (C) is the fourth (by mass) most abundant element in the universe. Due to its unique electronic structure, three types of chemical bonds, sp$^1$, sp$^2$, and sp$^3$ hybridizations, can be formed. Such a property facilitates C to build various multi-atomic structures with different molecular configurations, including amorphous carbon, diamonds, graphite, graphene, fullerenes, carbon buckyonions, carbon nanotubes, and planar polycyclic aromatic hydrocarbon (PAH) molecules \citep{henning1998carbon}.

Many of these carbonaceous compounds have been detected in the interstellar medium (ISM) either through their fingerprint spectral features in the infrared (IR) or through isotope analysis of primitive meteorites. The latter leads to the identification of presolar nanodiamonds and graphite of interstellar origin \citep{li2012nanodust}. Solid hydrogenated amorphous carbon reveals their presence in the low-density ISM through the 3.4$\mum$  aliphatic C--H absorption feature \citep{pendleton2002organic}. C$_{60}$ and C$_{70}$ as well as their cations are seen in space through their characteristic IR emission features, for instance, at 7.0, 8.45, 17.3 and 18.9$\mum$ for C$_{60}$ \citep{cami2010detection, sellgren2010c60, garcia2010formation, kwok2011mixed, berne2017detection} and at 6.4, 7.1, 8.2 and 10.5$\mum$ for C$_{60}^{+}$ \citep{berne2013, strelnikov2015observing}.

The distinctive set of broad emission bands at 3.3, 6.2, 7.7, 8.6, 11.3 and 12.7$\mum$, collectively known as the ``unidentified infrared'' emission (UIE) features ever since their first detection over four decades ago, are ubiquitously seen in a wide variety of astrophysical regions and are generally attributed to PAHs \citep{leger1984identification, allamandola1989interstellar}. The possible presence of planar C$_{24}$, a graphene sheet, in planetary nebulae of both our own galaxy and our nearest neighbour, the Magellanic Clouds, was revealed through the emission features at $\simali$6.6, 9.8, and 20$\mum$ \citep{garcia2011formation, garcia2012infrared}. \cite{chen2017graphene} found that $\simali$5$\ppm$ of C/H, i.e., $\simali$1.9\% of the total interstellar C could be locked up in the interstellar graphene. 

\begin{figure*}
\begin{center}
\includegraphics[width=0.5\textwidth]{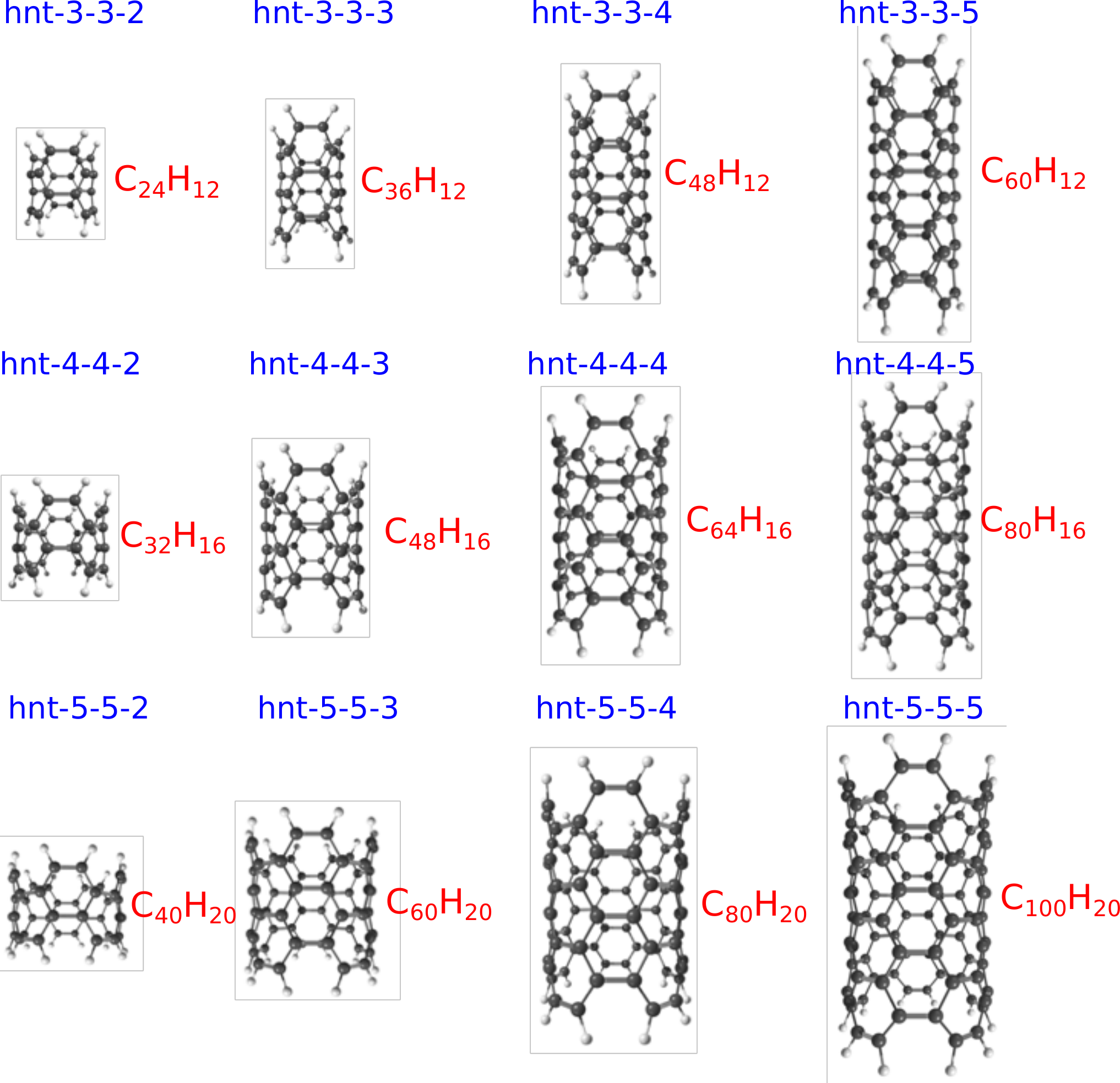}
\caption{\footnotesize
              The structures of the hydrogenated nanotubes (hnt) 
              studied in this work. The name of each nanotube is 
              shown on the top of each structure, 
              where the first two digits are the chiral index, 
              the last digit refers to the number of unit cells 
              of the nanotube. The chemical formulae of 
              the nanotubes are shown on the right hand 
              of each structure.
              }
\label{fig:nt}
\end{center}
\end{figure*} 

Interestingly, C$_{60}$ was initially synthesized in experiments aimed at understanding the formation of long-chain carbon molecules in the ISM. In the experiments, the remarkably stable cluster consisting of 60 carbon atoms are produced from graphite irradiated by laser \citep{kroto1985week}. More recently, both experimental and theoretical studies have shown that fullerenes and graphene flakes can be formed from large PAHs through dehydrogenation and isomerization \citep{berne2012, pietrucci2014fate, zhen2014}. In the first step, a large PAH molecule dissociates to a graphene sheet or graphene flake through dehydrogenation \citep{berne2012, zhen2014}. Then the graphene sheet/flake isomerizes to a cage structure, such as fullerene \citep{pietrucci2014fate}. In addition, the byproducts during fullerene formation, nanotubes, have first been described as helical microtubules of graphitic carbon in 1991 by Iijima, who generated the novel material by an arc discharge evaporation process originally designed for the production of fullerenes \citep{iijima1991helical}. Thereafter, many studies pointed out that carbon nanotubes could be considered as elongated fullerenes \citep{terrones2004shape, uberuaga2012fullerene, cruz2016fullerene}. Cruz-Silva et al.\ (2016) report that at an initial growth stage, single-walled carbon nanotubes begin to grow from a hemisphere-like fullerene cap. They find that the insertion of C$_2$ units convert the fullerene cap into an elongated structure that leads to the formation of very short carbon nanotubes. Using molecular dynamics methods, Uberuaga et al.\ (2012) simulated fullerene and graphene formation from carbon nanotubes. They find that small $(n,n)$ nanotubes with $n\simlt5$ quickly form closed structures, while the larger ones with $n\simgt6$ unfold to graphene sheets. As a stable isomer of graphene and fullerene, nanotubes could also be present in the ISM where both PAHs and fullerenes are present, through a feasible pathway described later.  

\section{Carbon Nanotubes: Nomenclature}\label{sec:cnt}
Carbon nanotubes belong to the family of synthetic carbon allotropes and are characterized by a network of sp$^2$ hybridized carbon atoms. They are conceivably constructed by rolling up a graphene sheet into a cylinder with the hexagonal rings joining seamlessly. Depending on the way the graphene sheet is rolled up, a huge diversity of single-walled carbon nanotube structures can be constructed, differing in length, diameter and roll-up angle, which defines the orientation of the hexagonal carbon rings in the honeycomb lattice relative to the axis of the nanotube. The (n,m) nanotube naming scheme can be thought of as a vector ($C_h$) in an infinite graphene sheet that describes how to ``roll up'' the graphene sheet to make the nanotube: 
\begin{equation} \label{eq:1}
C_h = n\mathbf{e_1} + m\mathbf{e_2}
\end{equation}
where \textbf{e$_1$} and \textbf{e$_2$} denote the unit vectors of graphene in real space. In order to describe the length of nanotubes, we introduce a third parameter $l$, to characterize the number of unit cells in the nanotubes. Nanotubes can be rather long, with thousands of unit cells, i.e., $l > 10^{3}$. However, it is unrealistic to run {\it ab initio} calculations for such huge species, therefore we focus our study on nanotubes (NTs) with limited diameters ($n\simlt5$, $m\simlt5$) and lengths $l\simlt5$. To investigate the C--H stretching bans, i.e., the band at $\simali$3.3$\mum$, hydrogenated nanotubes (HNTs) are investigated. In the hydrogen--rich interstellar environments, NTs are expected to be hydrogenated. We use the abbreviation hnt-n-m-l as the unique identifier for hydrogenated nanotubes. Figure~\ref{fig:nt} shows some examples of the optimized structures of HNTs studied in this work. 

\section{Computational Method}
In this work, the ab initio calculations are carried out using density functional theory (DFT). The dissociation energies and transition state energies are computed using the hybrid density functional B3LYP \citep{becke,lee1988development} as implemented in the Gaussian 16 program \citep{frisch2016gaussian}. All structures are optimized using the 6-311++G(2d,p) basis set. It has been reported that such combination of functional and basis sets, i.e., B3LYP/6-311++G(2d,p), produces reasonable ionization, dissociation and transition state energies for PAHs and the calculated results are consistent with high accuracy method, e.g., CBS-QB3 method \citep{holm2011,johansson2011unimolecular,chen2015}. To take the intermolecular forces into account, the D3 version of Grimme's dispersion with Becke--Johnson damping \citep{grimme2011effect} is included in the calculations.

The vibrational frequencies are calculated for the optimized geometries to verify that these correspond to minima or first-order saddle points (transition states) on the potential energy surface (PES). The zero point vibrational energies (ZPVE) are taken into account. Intrinsic reaction coordinate (IRC) calculations \citep{fukui,dykstra} are performed to confirm that the transition state structures are connected to their corresponding local PES minima. We focus on neutral and cation systems with the lowest multiplicities and only the ground state PES are considered, as the non-radiative decay for such a large molecule is very rapid: $\simali$10$^{-12}\s$ \citep{vierheilig1999femtosecond, zewail2000femtochemistry}.

The harmonic IR spectra are calculated with a smaller basis set, viz. B3LYP/6-31+G(d), based on the optimized structures obtained at the same level of theory. Such a combination produces accurate vibrational spectra --- previous studies have shown that larger basis sets, could not improve the spectra significantly \citep{merrick2007evaluation}. To achieve more accurate IR spectra, anharmonic effects should be taken into account, which would appreciably affect the number of bands, band positions and relative intensities \citep{mackie2015anharmonic,chen2018carrier}.

\section{Synthesizing Carbon Nanotubes through Hydrogen-Abstraction and Acetylene-Addition}\label{sec:haca}
The hydrogen-abstraction/acetylene-addition (HACA) has been suggested as a formation mechanism of PAHs in the outflows of carbon-rich asymptotic giant branch (AGB) stars \citep{frenklach1985detailed, frenklach1989, richter2000formation, frenklach2002reaction}. In the HACA mechanism, a repetitive hydrogen loss from the aromatic hydrocarbon is followed by addition of one or two acetylene molecule(s) before cyclization and aromatization \citep{frenklach1985detailed, frenklach1989, richter2000formation, appel2000kinetic}. Using HACA, \cite{parker2014hydrogen} demonstrated in crossed molecular beam experiments combined with electronic structure and quantum statistical calculations that naphthalene (C$_{10}$H$_8$) can be synthesized via successive reactions of the phenyl radical with two acetylene molecules. \cite{yang2017haca} showed that phenanthrene (C$_{14}$H$_{10}$) can also be formed by the reaction of the biphenylyl radical (C$_{12}$H$_9$\text{\textbullet}) with a single acetylene molecule through addition to the radical site followed by cyclization and aromatization. Very recently, \cite{zhao2018pyrene} found a facile, isomer--selective formation pathway of pyrene (C$_{16}$H$_{10}$) using the HACA method, and they suggested that molecular mass growth from pyrene may lead through systematic ring expansions not only to more complex PAHs, but ultimately to 2D graphene--type structures. However, the validity of HACA to form 3D nanostructures has remained uncharted. 

\begin{figure*}
\begin{center}
\includegraphics[width=0.8\textwidth]{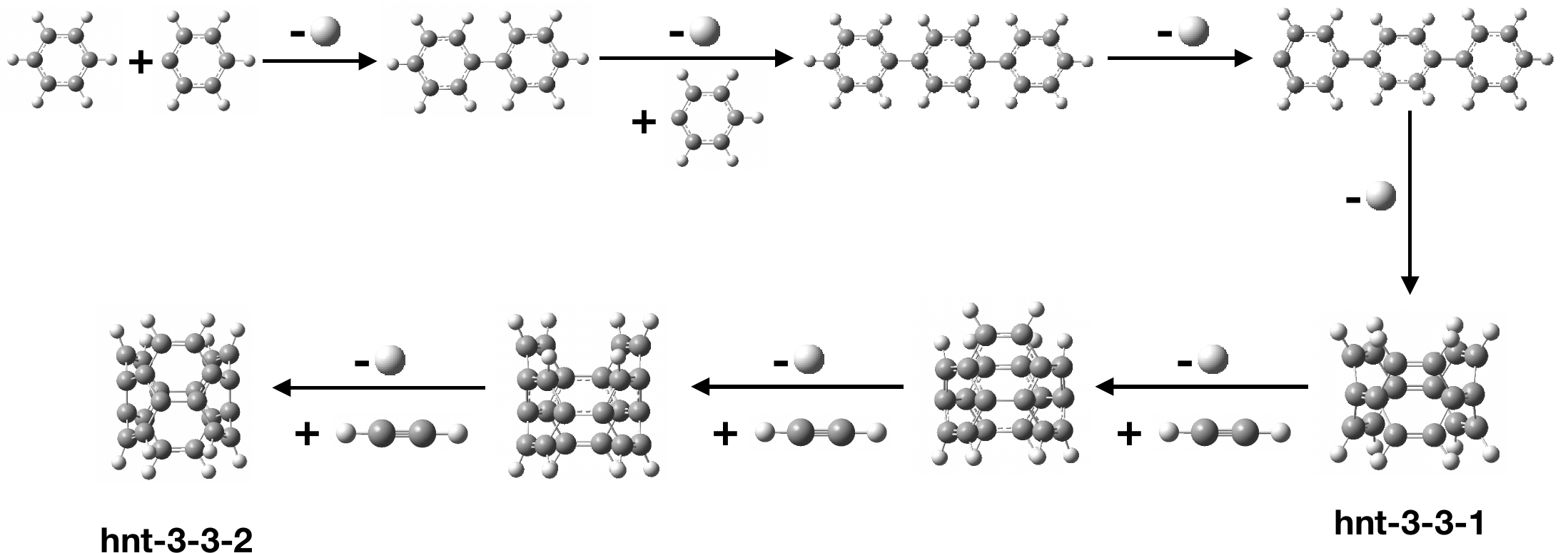}
\caption{\footnotesize
              A molecular growth process of 
              3D nanostructures involving HACA.}
\label{fig:general}
\end{center}
\end{figure*}

Figure~\ref{fig:general} shows the molecular mass growth process from planar to 3D structures involving HACA: at the beginning, a benzene combines with a phenyl to form a biphenyl through hydrogen abstraction. Then, the biphenyl loses a hydrogen atom and reacts with another phenyl to form a triphenyl. Following a H--loss, the triphenyl isomerizes to a closed 3D structure, hnt-3-3-1. The rest growth processes to more complex hydrogenated nanotubes are similar to the HACA processes for PAHs \citep{yang2017haca, zhao2018pyrene}.

\begin{figure*}
\begin{center}
\includegraphics[width=0.85\textwidth]{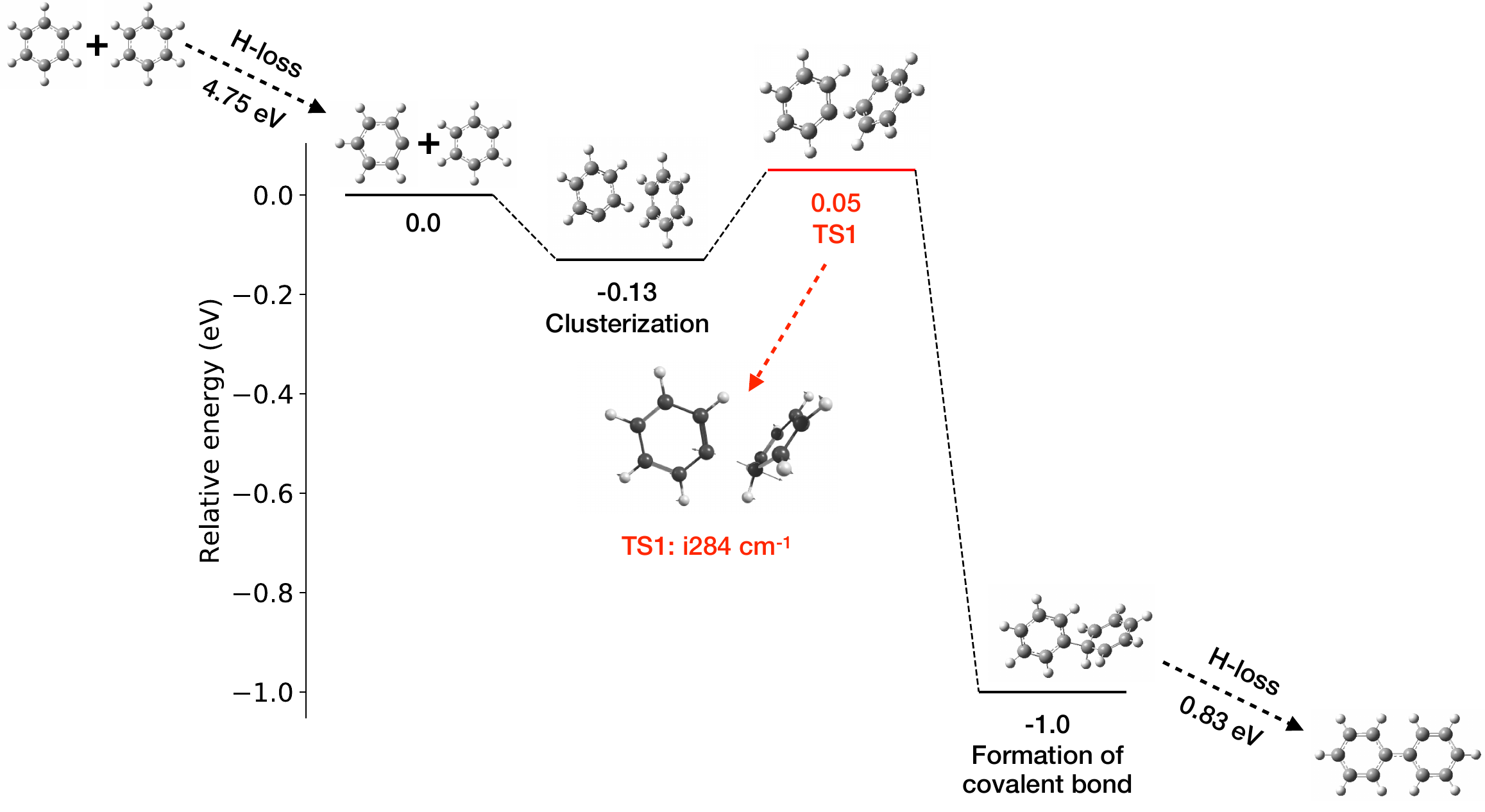}
\caption{\footnotesize
               Calculated reaction barriers and dissociation energies 
               for the formation of biphenyl from two benzene molecules. 
               The energies are calculated based on the left system (phenyl + benzene).
               The arrow indicates the intensities of the imaginary frequencies and 
               displacement vectors of the transition state.}
\label{fig:ts1}
\end{center}
\end{figure*}

\begin{figure*}
\begin{center}
\includegraphics[width=0.8\textwidth]{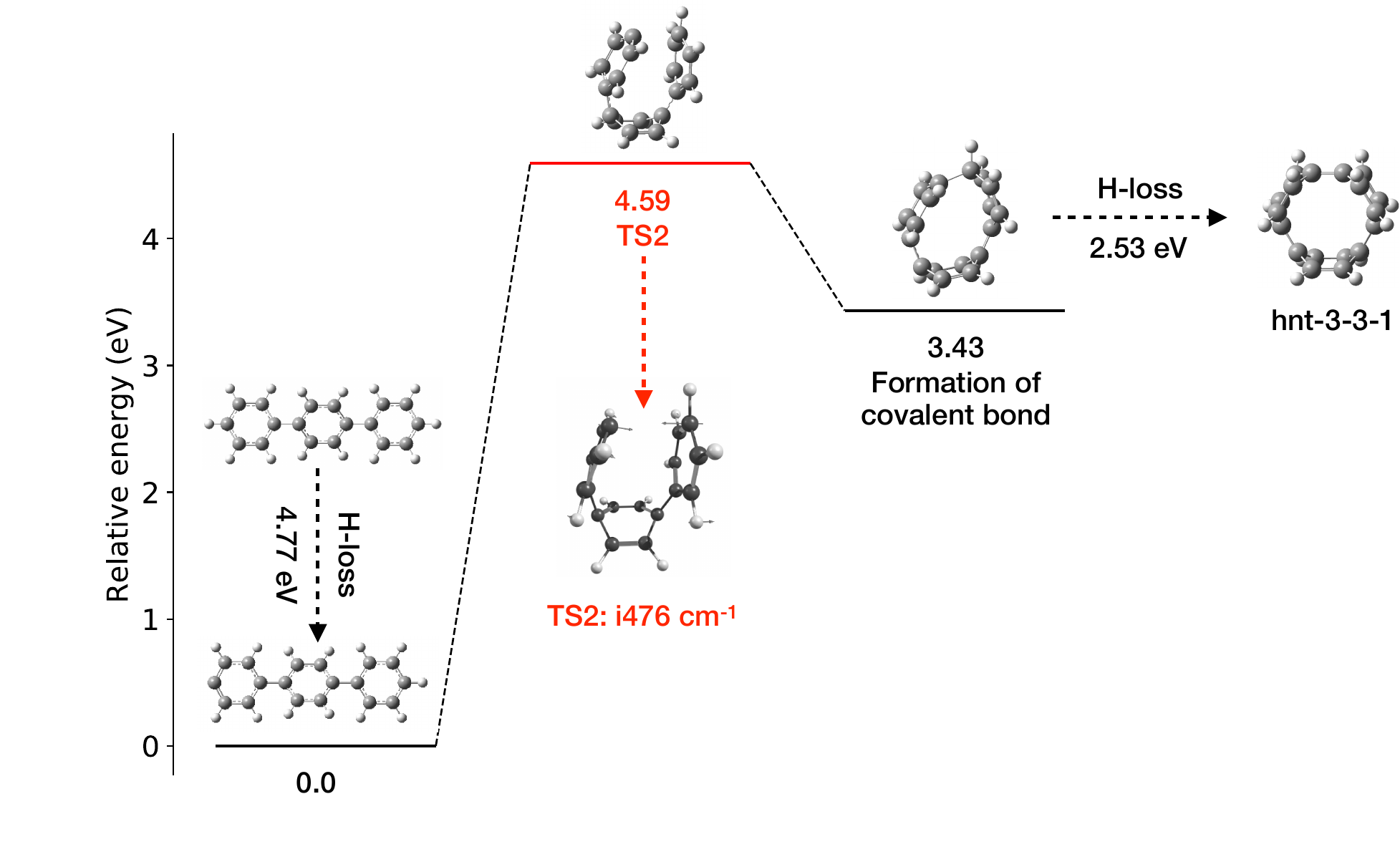}
\caption{\footnotesize
               Calculated reaction barrier for the molecular bending.
               The energies are calculated based on the left molecule (terphenyl-H).
               The arrow indicates the intensity of the imaginary frequency and 
               displacement vector of the transition state.}
\label{fig:ts2}
\end{center}
\end{figure*}

\begin{figure*}
\begin{center}
\includegraphics[width=0.85\textwidth]{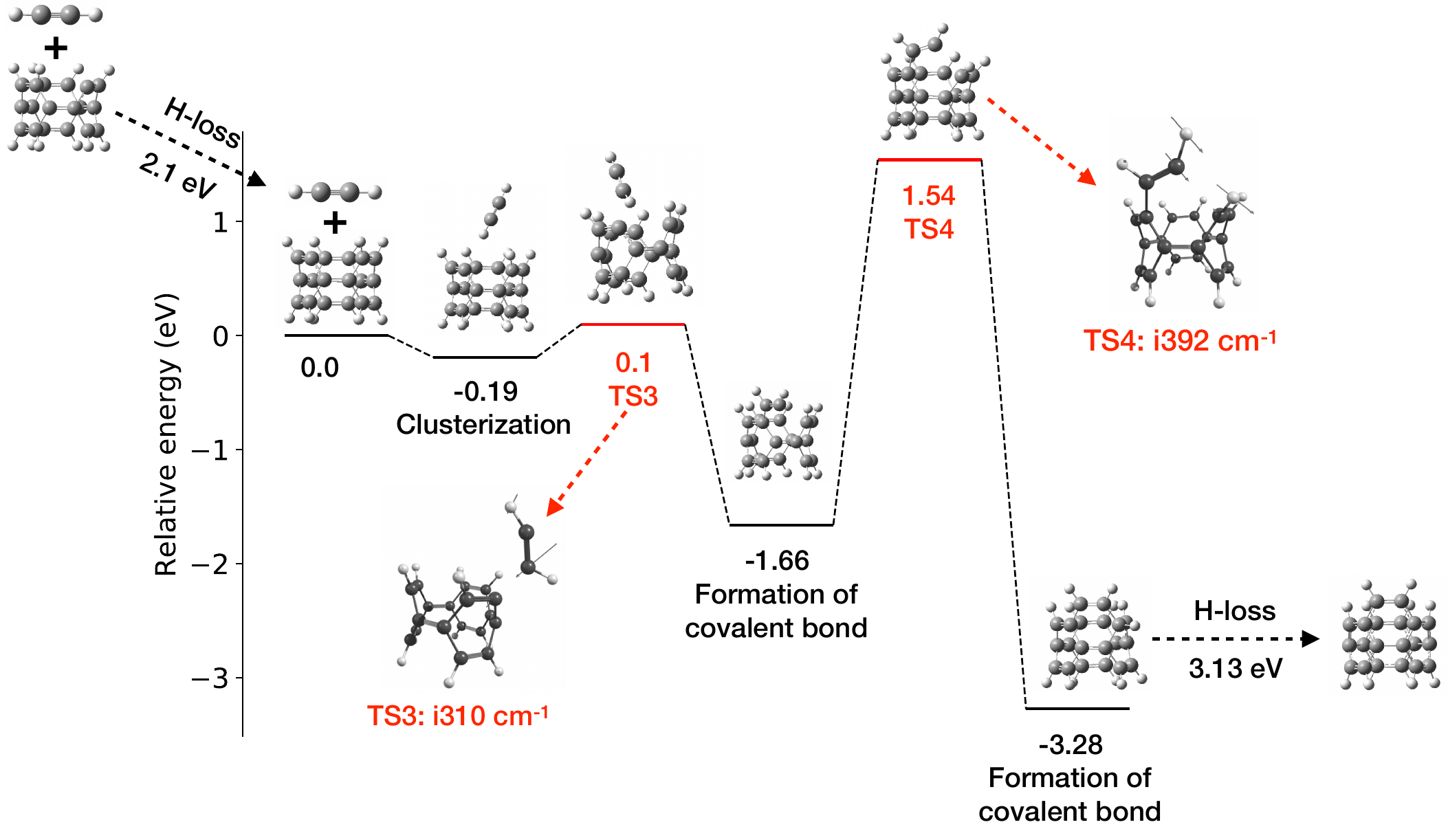}
\caption{\footnotesize
               Calculated reaction barriers and dissociation energies 
               for the molecular growth process in the 3D nanostructure (hnt-3-3-1). 
               The energies are calculated based on the left system. The arrows indicate
               the intensities of the imaginary
               frequencies and displacement vectors of the corresponding transition 
               states.}
\label{fig:ts3}
\end{center}
\end{figure*}

Multiple barriers exist in the proposed formation pathway shown Figure~\ref{fig:general}. Using DFT, we have studied the key barriers (transition states) along the formation pathway. Figure~\ref{fig:ts1} shows the formation of biphenyl from two neutral benzenes. The calculated electronic, zero point vibrational and total energies of the main strutures are given in Appendix. The reaction pathway begins with the loss of a hydrogen atom from a benzene molecule. A phenyl molecule will be formed following the H-loss from benzene which could clusterize with a neighbor benzene through van der Waals forces. The binding energy of the weakly bonded cluster is found to be 0.13 $\eV$. After that, a 0.18 $\eV$ barrier (shown as TS1 in Figure~\ref{fig:ts1}) has to be overcame to form a covalent bond between the phenyl and benzene molecules. To prove the nature of the transition state, we exhibit the imaginary frequency and displacement vectors of the transition state. As shown in Figure~\ref{fig:ts1}, the intensity of the imaginary frequency is about {\it i}284 cm$^{-1}$. The displacement vectors show that the ``bald'' carbon (without attached hydrogen) on the phenyl side interacts with a carbon on the benzene side, which clearly indicate the tendency to form a covalent bond in between the two carbons. In order to confirm the transition state calculations, we have also investigated the intrinsic reaction coordinates (IRC) transition state, see Appendix for details. In addition, both experimental and theoretical studies have confirmed that the formation of covalent bond in PAH or fullerene clusters could occur efficiently in the interstellar conditions \citep{zettergren2013formations, zhen2018laboratory, chen2018formation}. 

A biphenyl could be formed following the loss of the ``extra'' hydrogen in between phenyl and benzene (see Figure~\ref{fig:ts1} for the location of the hydrogen). This process can be repeated to form terphenyl or other larger 2D nanostructures \citep{yang2017haca, zhao2018pyrene}. However, it is unclear how the closed 3D nanostructures could be produced. We find that the molecular bending or curvature is crucial for the formation of 3D geometries \citep{chen2017planes}. Moreover, it has been shown that linear molecules do bend significantly at vibrationally--excited states without breaking the bonds \citep{chen2019formation}. Figure~\ref{fig:ts2} shows the key transition state for the molecule bending, in which a terphenyl convert to a closed 3D nanostructure following the loss of two hydrogens. The imaginary frequency for such transition state is about {\it i}476$\cm^{-1}$. The displacement vectors show that the atoms on both edges of the molecule wag simultaneously towards or outwards each other, which demonstrate the tendency to close the structure. Figure~\ref{fig:irc2} in the Appendix shows the IRC of the barrier for molecular bending (TS2 in Figure~\ref{fig:ts2}). The highest point of the IRC represents the transition state, the left side is the closed nanostructure, while the right side corresponds to the open one (after optimization). One can see that this IRC provides reasonable reaction pathway between the reactant and the product. 

Starting from the closed 3D nanostructure (e.g., hnt-3-3-1), The acetylene addition on hydrogenated nanotubes takes about four steps: (i) H-loss from the edge of the HNT, (ii) an acetylene overcomes a barrier (TS3 in Figure~\ref{fig:ts3}) to form a covalent bond with the HNT, (iii) it goes through another barrier (TS4 in Figure~\ref{fig:ts3}) to form a closed hexagon structure, (iv) the dissociation of the ``extra'' hydrogen (see Figure~\ref{fig:ts3} for the location of the hydrogen). Following these four steps, the molecule can grow longer and more complex. 

\section{Infrared Vibrational Sepctra of Carbon Nanotubes}\label{sec:ir}

\begin{figure*}
\begin{center}
\includegraphics[width=0.8\textwidth]{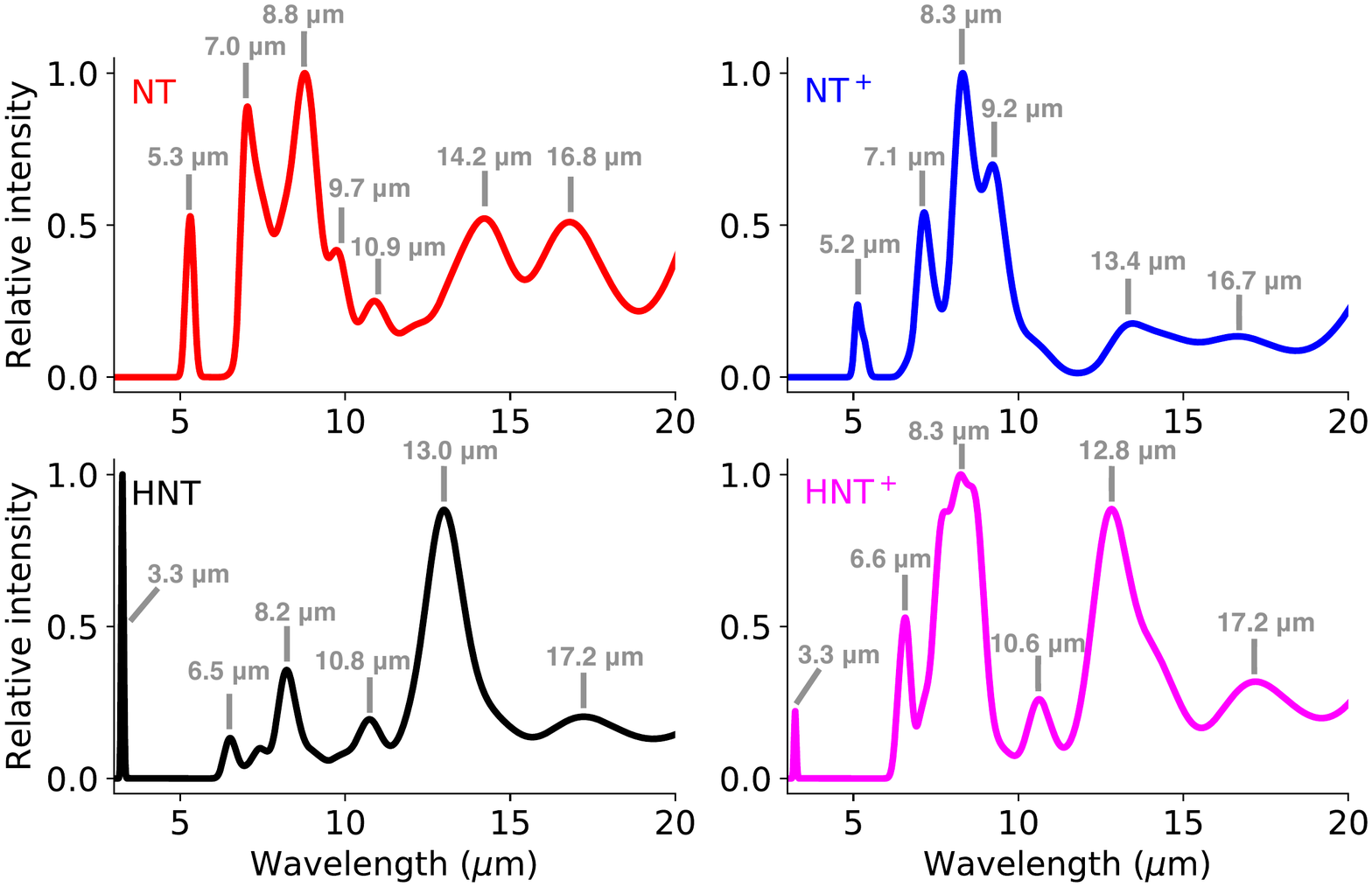}
\caption{\footnotesize
              DFT-computed IR vibrational spectra (convolved with Gaussian profiles of FWHM of 30$\cm^{-1}$) of NTs, HNTs and their cations. 
              }
\label{fig:ir}
\end{center}
\end{figure*} 

Figure~\ref{fig:ir} shows the DFT-calculated spectra of NTs, HNTs and their cations in the spectral range of 3--20$\mum$. The typical IR bands are also listed in Table~\ref{tab:comparison} and compared with other carbon compounds of astrophysical interest, including PAHs, fullerene, and graphene sheet. For each transition, we assign a full width at half maximum (FWHM) of 30$\cm^{-1}$ which is consistent with the natural line width expected from free--flying molecule \citep{allamandola1989interstellar, bauschlicher2008infrared, boersma2011polycyclic, ricca2012infrared}. This natural line width arises from intramolecular vibrational energy redistribution. 


The molecules included in Figure~\ref{fig:ir} are nt-3-3-2, nt-3-3-3, nt-3-3-4, nt-3-3-5, nt-4-4-2, nt-4-4-3, nt-4-4-4, nt-4-4-5, nt-5-5-2, nt-5-5-3, nt-5-5-4, and nt-5-5-5 as well as their cations. Also included in Figure~\ref{fig:ir} are the hydrogenated counterparts of these neutral and cationic NTs. For these 48 spectra, the vibrational bands mostly lay at about 3.3, 5.3, 6.0--9.4, 10.6, 12--15 and 16--18$\mum$. We obtain the mean spectra of NTs and HNTs by averaging the individual spectrum of each species on a per carbon atom basis and find that the dominant bands are at $\simali$5.3, 7.0, 8.8, 9.7, 10.9, 14.2, and 16.8$\mum$ for NTs and at $\simali$3.3, 6.5, 8.2, 10.8, 13 and 17.2$\mum$ for HNTs, respectively. For their cationic counterparts, the major bands occur at $\simali$5.2, 7.1, 8.3, 9.2, 13.4 and 16.7$\mum$ for NT cations and at $\simali$3.3, 6.6, 8.3, 10.6, 12.8 and 17.2$\mum$ for HNT cations, respectively. Most noticeably, the 5.3$\mum$ band associated with the C=C symmetric stretching modes on the edge of NTs is absent in HNTs and other carbon compounds of astrophysical interest (see Table~\ref{tab:comparison} for details), while HNTs exhibit a prominent C--H stretching band at 3.3$\mum$ and a prominent C--H out-of-plane bending band at 12.8--13$\mum$. The later two bands (3.3 and 12.8--13$\mum$) are rather similar to the UIE bands (3.3 and 12.7$\mum$) which are commonly attributed to the C--H stretching and C--H out-of-plane bending of PAHs. In addition, the C--C stretching bands at 6.5--6.6$\mum$ and 8.2--8.3$\mum$ of HNTs are also close to the 6.2, 7.7 and 8.6$\mum$ UIE bands. Although some of these bands are difficult to disentangle from the UIE bands, the unprecedented sensitivity of the {\it James Webb Space Telescope} (JWST) scheduled to be launched in 2021 will place the detection (or non-detection) of the bands at $\simali$5.3, 6.5--6.6, 7--7.1, 8.2--8.3, 10.6--10.8, 12.8--13.4, 14.2 and 16.8--17.2$\mum$ on firm ground and enable far more detailed band analysis and formation pathway than previously possible. The theoretical IR spectra of both neutral and cationic NTs and their hydrogenated counterparts could also be used for possible searches of these species in the ISM with exisiting observations made by, e.g., the {\it Infrared Space Observatory} and the {\it Spitzer Space Telescope}.

\begin{table*}
\begin{center}
\caption{Dominant IR features of carbon compounds of astrophysical interest.}
\begin{tabular}{l|l}
\hline
\hline
Molecule & Dominant IR bands\\
&\\
\hline
Fullerene (C$_{60}$) & 7.0, 8.45, 17.3, 18.9$\mum$ \\
&\\
Fullerene cation (C$_{60}^{+}$) & 6.4, 7.1, 8.2, 10.5$\mum$ \\
&\\
Graphene sheet (C$_{24}$) & 6.6, 9.8, 20$\mum$\\
&\\
PAHs & 3.3, 6.2, 7.7, 8.6, 11.3, 12.7$\mum$ \\
&\\
Nanotubes (NT) & 5.3, 7.0, 8.8, 9.7, 10.9, 14.2, 16.8$\mum$ \\
&\\
Nanotube Cations (NT$^+$) & 5.2, 7.1, 8.3, 9.2, 13.4, 16.7$\mum$ \\
&\\
Hydrogenated nanotubes (HNT) & 3.3, 6.5, 8.2, 10.8, 13, 17.2$\mum$ \\
&\\
Hydrogenated nanotube cations (HNT$^+$) & 3.3, 6.6, 8.3, 10.6, 12.8, 17.2$\mum$ \\

\hline 
\hline
\end{tabular}
\label{tab:comparison}
\end{center}
\end{table*}

\section{Summary}
Using DFT, we have demonstrated that CNTs could form in the ISM through the hydrogen abstraction and acetylene addition mechanism. CNTs and hydrogenated CNTs as well as their cationic counterparts exhibit intense vibrational transitions in the IR. The possible presence of these species in space will be tested by the upcoming {\it JWST}. 

\section*{Acknowledgements}
We thank the anonymous referee for his/her very helpful comments which considerably improved the presentation of this work. We also thank Shaoqi Zhan for sharing his insights on transition state calculations. This work is supported by the Swedish Research Council (Contract No. 2015--06501). The calculations were performed on resources provided by the Swedish National Infrastructure for Computing (SNIC) at the High Performance Computing Center North (HPC2N). A.L. is supported in part by NSF AST--1816411 and NASA 80NSSC19K0572.

\newpage

\chead{\Large{\textbf{Appendix}}}

\begin{figure}
\begin{center}
\includegraphics[width=1.0\columnwidth]{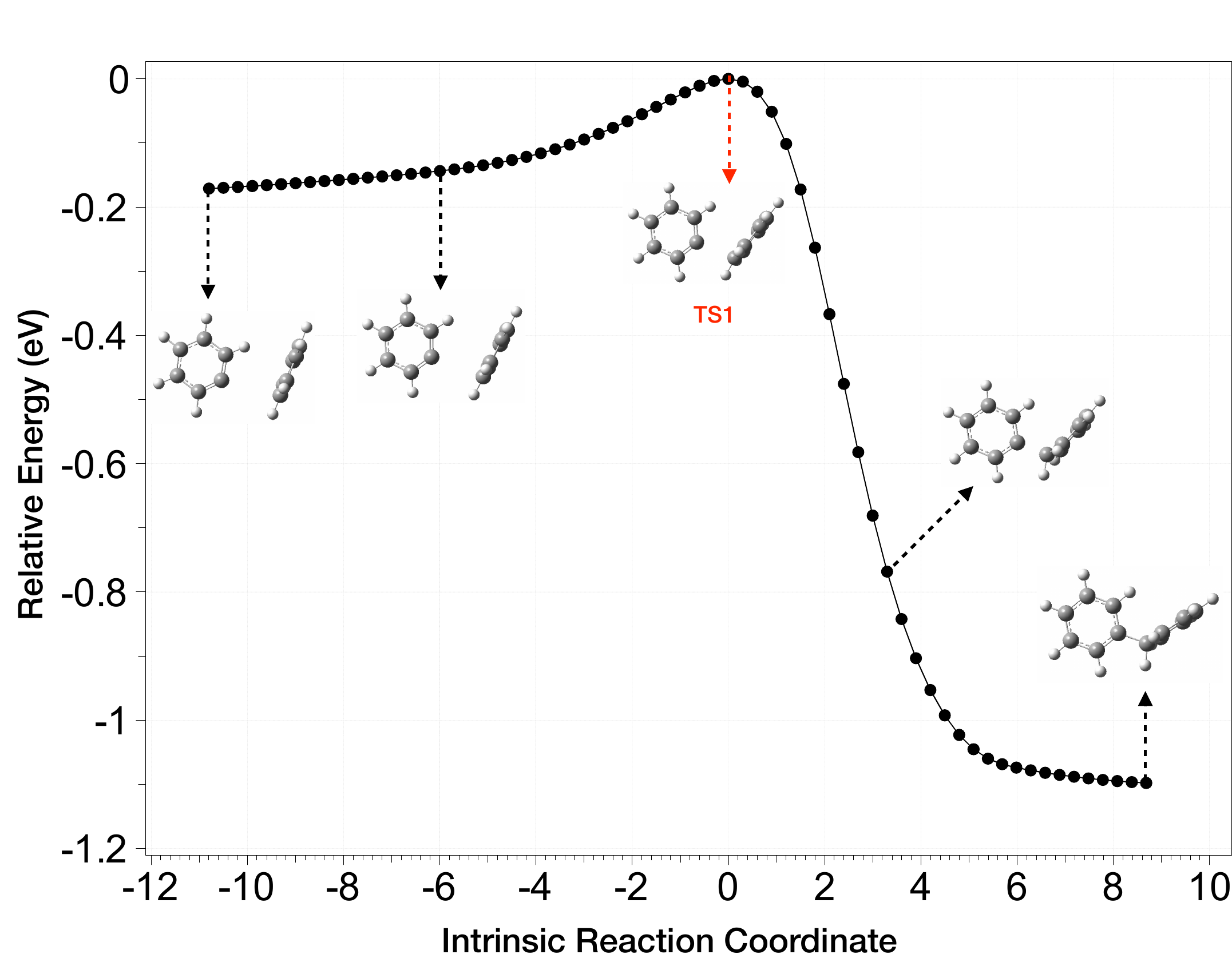}
\caption{\footnotesize
               The intrinsic reaction coordinates (IRC) for the transition state of 
               TS1 (see Figure~\ref{fig:ts1}). The arrows indicate
               the structures at the corresponding IRC steps.}
\label{fig:irc1}
\end{center}
\end{figure}

\begin{figure}
\begin{center}
\includegraphics[width=1.0\columnwidth]{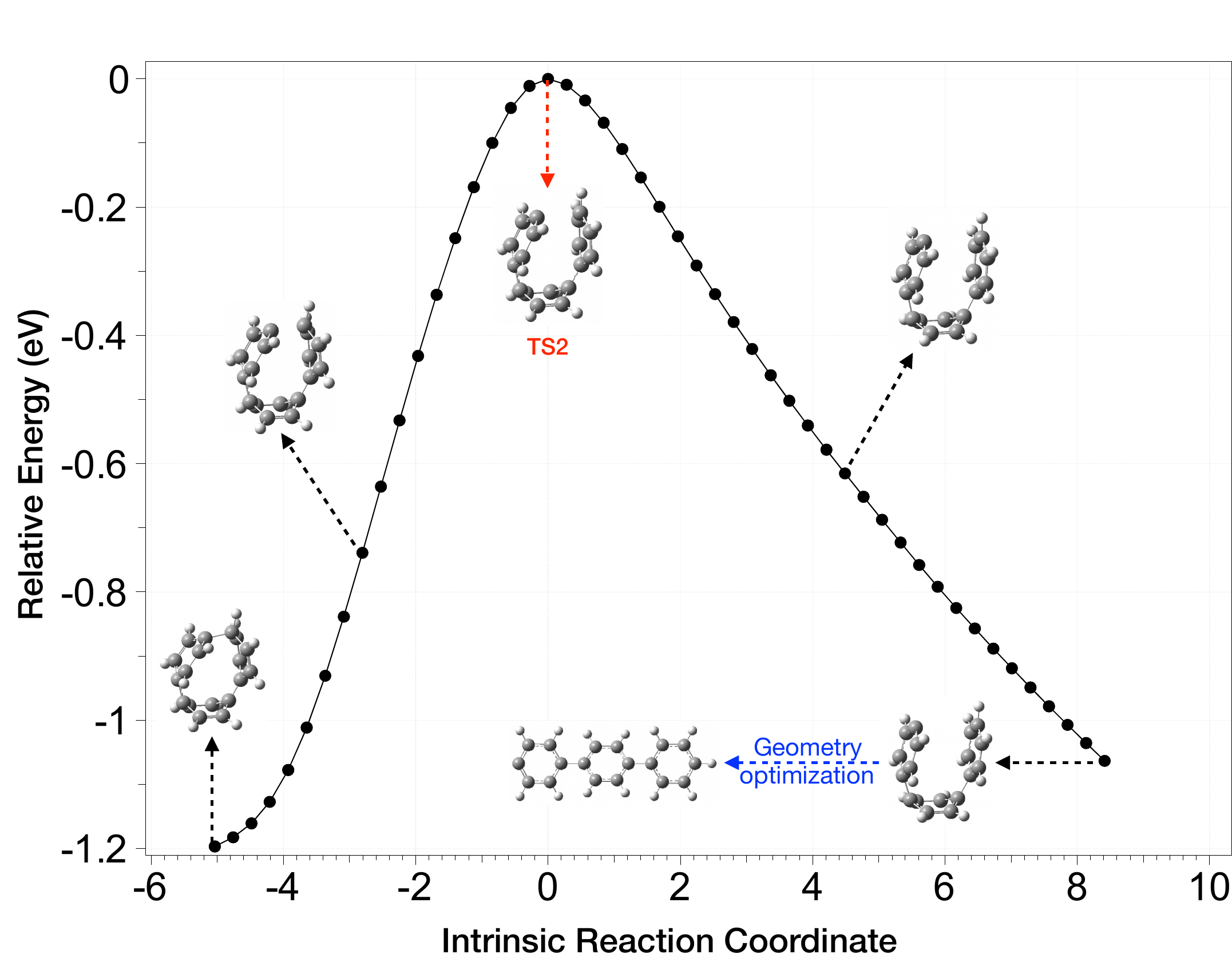}
\caption{\footnotesize
               The IRC for the transition state of 
               TS2 (see Figure~\ref{fig:ts2}). The structure on the
               right--bottom corner represents the optimized geometry of the right--end 
               IRC step.}
\label{fig:irc2}
\end{center}
\end{figure}

\begin{figure}
\begin{center}
\includegraphics[width=1.0\columnwidth]{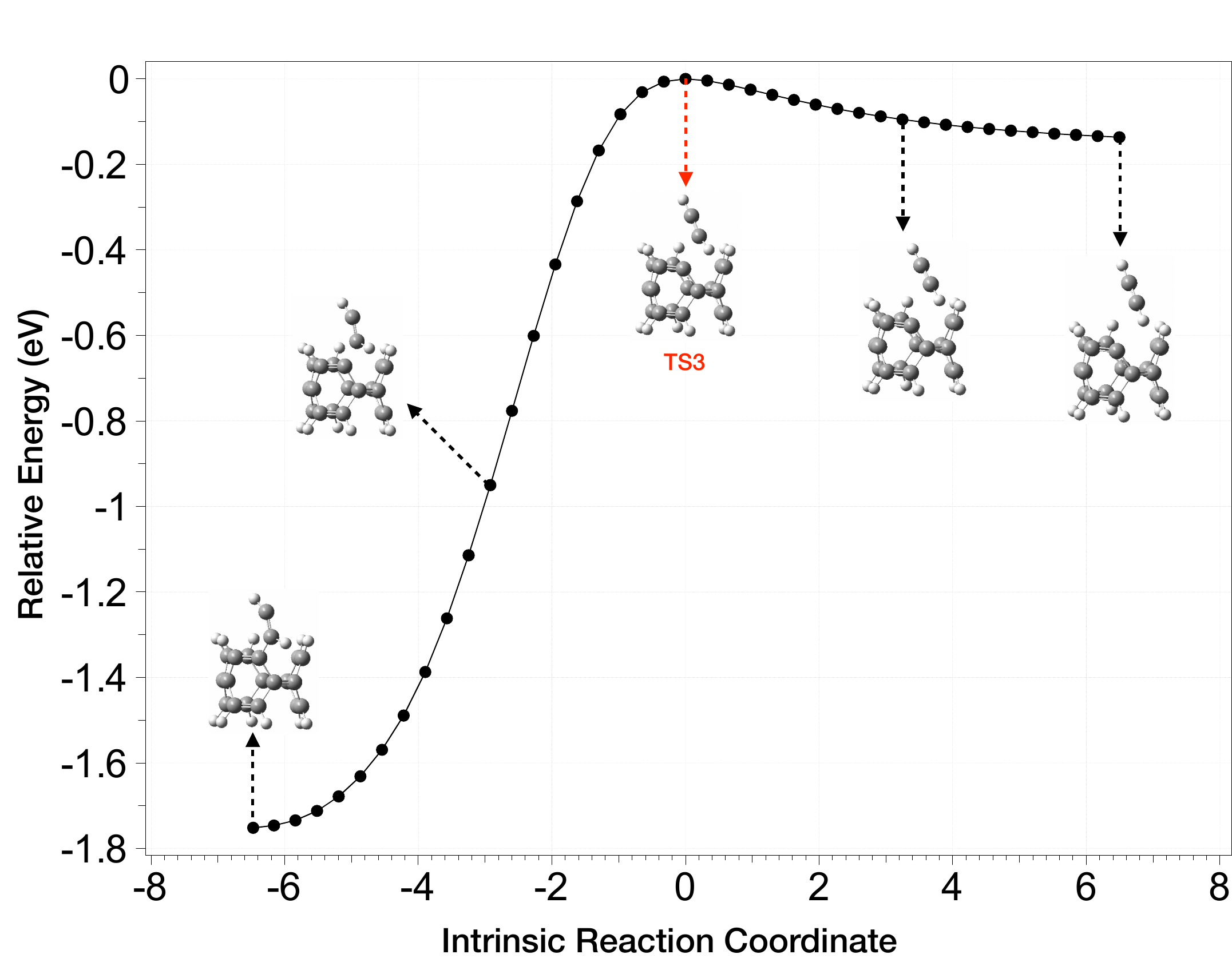}
\caption{\footnotesize
               The IRC for the transition state of 
               TS3 (see Figure~\ref{fig:ts3}). }
\label{fig:irc3}
\end{center}
\end{figure}

\begin{figure}
\begin{center}
\includegraphics[width=1.0\columnwidth]{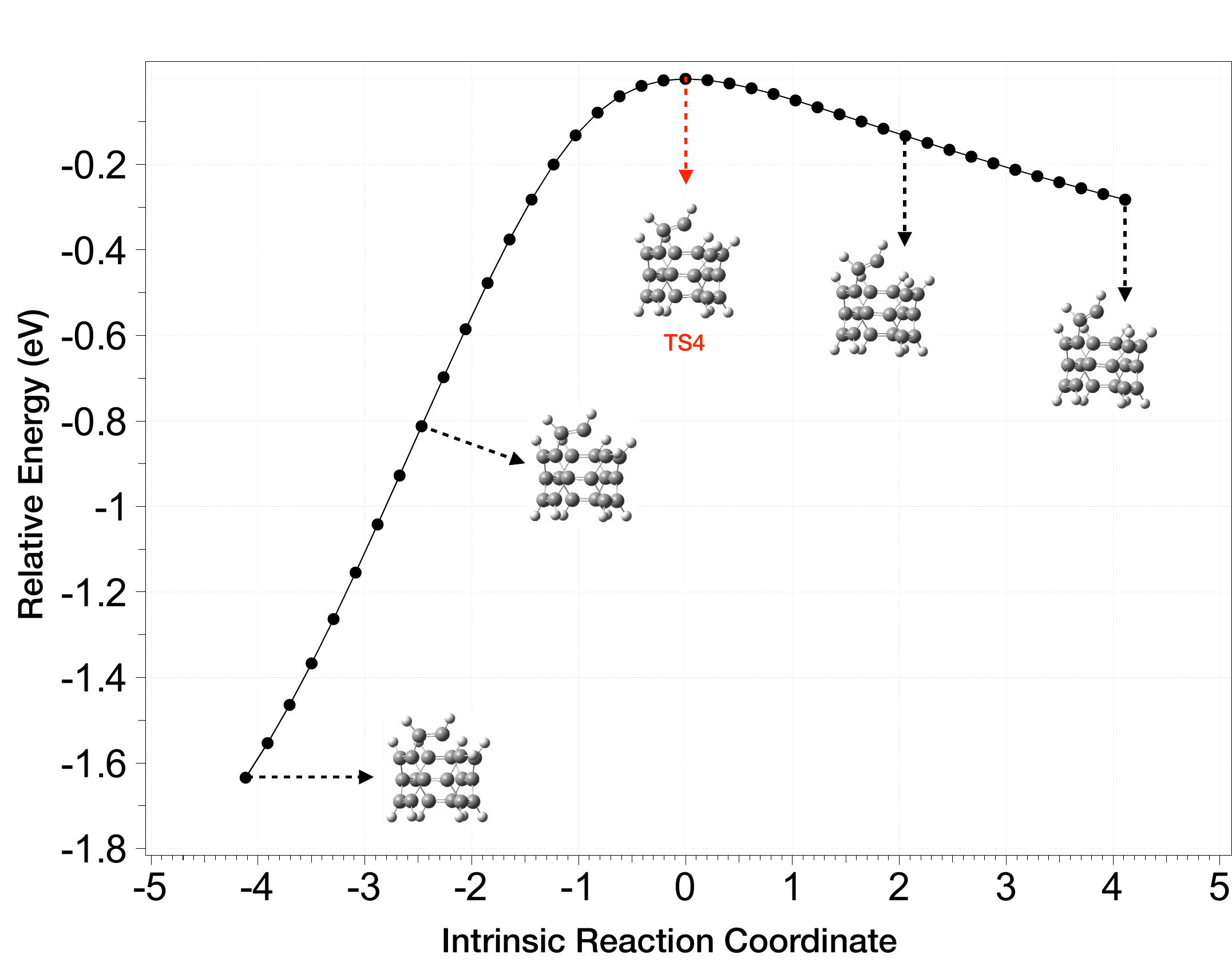}
\caption{\footnotesize
               The IRC for the transition state of 
               TS4 (see Figure~\ref{fig:ts3}). }
\label{fig:irc4}
\end{center}
\end{figure}

\begin{figure*}
\hspace{-3cm}\includegraphics[width=1.4\textwidth]{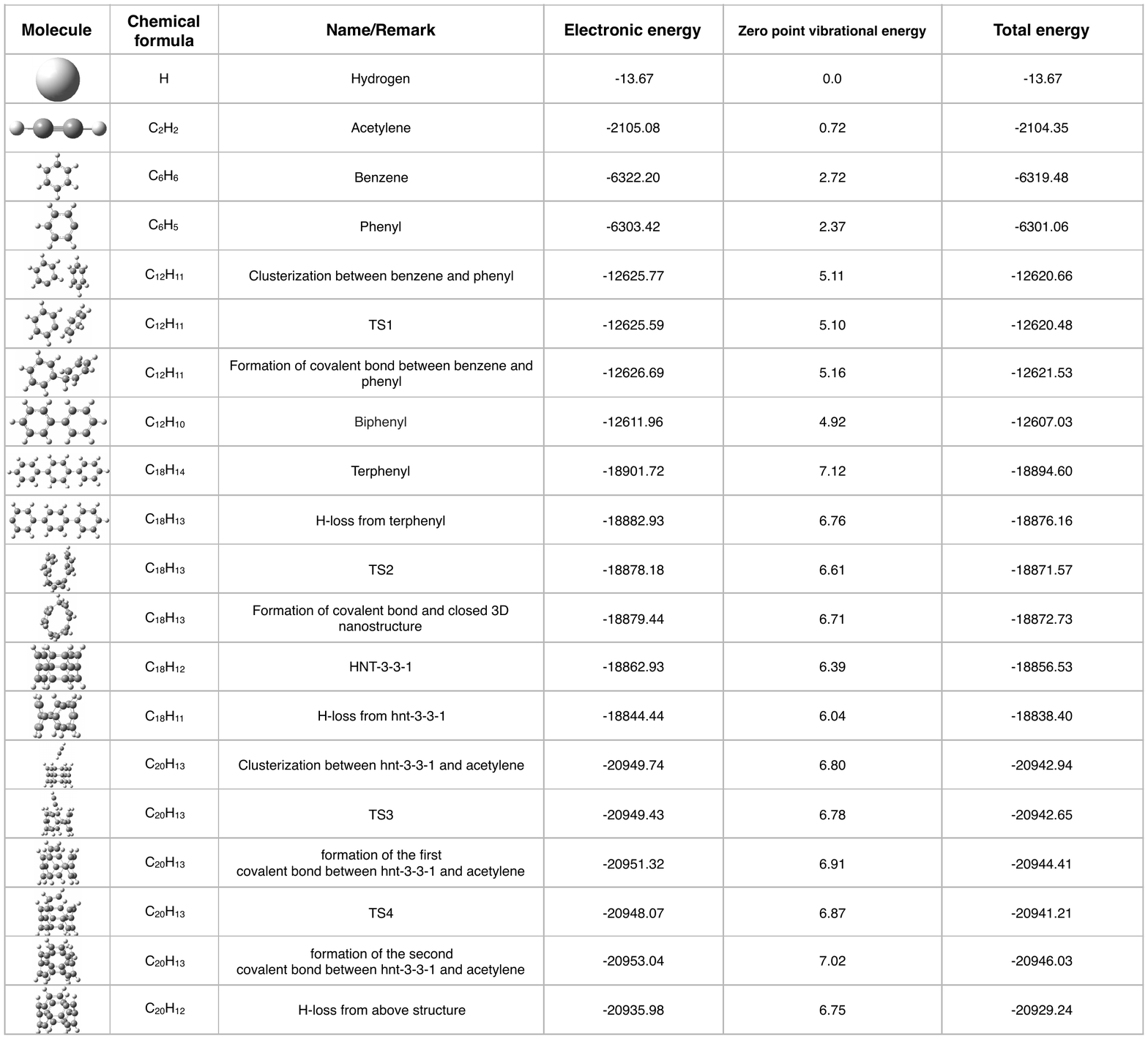}
\caption{\footnotesize
               DFT calculations (B3LYP/6-311++G(2d,p)) of electronic, zero point vibrational and total energies of the studied molecules. The values are given in eV.}
\label{fig:e}
\end{figure*}

\end{document}